\documentclass[aps,twocolumn,superscriptaddress,showpacs]{revtex4}

\usepackage{epsfig}
\usepackage{amsmath}
\usepackage{times}

\begin{document}

\title{The development of generalized synchronization on complex networks}

\author{Shuguang Guan}
\affiliation{Temasek Laboratories,
National University of Singapore, Singapore 117508}
\affiliation{Beijing-Hong Kong-Singapore
Joint Center of Nonlinear and Complex systems (Singapore),
Singapore 117508}

\author{Xingang Wang}
\affiliation{Temasek Laboratories, National University of
Singapore, Singapore 117508}
\affiliation{Beijing-Hong
Kong-Singapore Joint Center of Nonlinear and Complex systems
(Singapore), Singapore 117508}

\author{Xiaofeng Gong}
\affiliation{Temasek Laboratories, National University of
Singapore, Singapore 117508}
\affiliation{Beijing-Hong
Kong-Singapore Joint Center of Nonlinear and Complex systems
(Singapore), Singapore 117508}

\author{Kun Li}
\affiliation{Temasek Laboratories, National University of
Singapore, Singapore 117508}
\affiliation{Beijing-Hong
Kong-Singapore Joint Center of Nonlinear and Complex systems
(Singapore), Singapore 117508}

\author{C.-H. Lai}
\affiliation{Beijing-Hong Kong-Singapore
Joint Center of Nonlinear and Complex systems (Singapore),
Singapore 117508}
\affiliation{Department of Physics, National University of Singapore,
Singapore 117543}

\date{\today}

\begin{abstract}

In this paper, we investigate the development of generalized
synchronization (GS) on typical complex networks, such as
scale-free networks, small-world networks, random networks and
modular networks. By adopting the auxiliary-system approach to
networks, we show that GS can take place in oscillator networks
with both heterogeneous and homogeneous degree distribution,
regardless of whether the coupled chaotic oscillators are
identical or nonidentical. For coupled identical oscillators on
networks, we find that there exists a general bifurcation path
from initial non-synchronization to final global complete
synchronization (CS) via GS as the coupling strength is increased.
For coupled nonidentical oscillators on networks, we further
reveal how network topology competes with the local dynamics to
dominate the development of GS on networks. Especially, we analyze
how different coupling strategies affect the development of GS on
complex networks. Our findings provide a further understanding for
the occurrence and development of collective behavior in complex
networks.

\end{abstract}

\maketitle

\section{Introduction}

Synchronization in coupled chaotic oscillators has been
extensively studied in the past twenty years
\cite{Chaos_syn_reviews}, including complete synchronization
\cite{CS}, generalized synchronization
\cite{GS,GLL:2006,ARS:1996,ZWC:2002,GKL:2006}, and phase
synchronization (PS) \cite{PS}, etc. In CS, the dynamics of two
coupled systems totally coincide with each other; in GS, certain
functional relation exists between the dynamics of two coupled
systems which are usually nonidentical. Moreover, PS is a weaker
synchronization form in which the phases of two oscillators can be
locked while their amplitudes remain uncorrelated and chaotic.
Recently, the study of synchronization has been extended to the
area of complex networks
\cite{WS:1998,SW,BP:2002,HCK:2002,BA:1999,SF,MM:2005,Modular,PLGK:2006,ZZC:2007,GWLL:2008,Weighted,WLL:2007,ROH:2005,GMA:2007}.
For example, synchronization on small-world networks
\cite{WS:1998,SW,BP:2002,HCK:2002}, scale-free networks
\cite{BA:1999,SF,MM:2005}, modular networks
\cite{Modular,PLGK:2006,ZZC:2007,GWLL:2008},weighted networks
\cite{Weighted}, and gradient networks \cite{WLL:2007}, have been
investigated. These studies aim to explore the interplay between
network topology and dynamics on network. They are important for
us to understand the real situations in complex systems comprising
interacting elements in nature and human society.

In literatures, there are two main approaches studying
synchronization in complex networks. One approach is to study
synchronization of phase oscillators on complex networks, i.e.,
the generalized Kuramoto models
\cite{HCK:2002,ROH:2005,GMA:2007,GWLL:2008,Kuramoto:1979}. In this
model, the node dynamics is very simple which is governed by an
ideal phase oscillators, $\dot{\phi}=\omega$, where $\omega$ is
the frequency. The heterogeneity in the node dynamics can be
modeled by assigning different, usually random, frequencies to
different phase oscillators. The generalized Kuramoto models in
complex networks have the advantage that they can still be treated
analytically in many aspects \cite{ROH:2005,GWLL:2008}. The second
approach is to study the synchronizability of complex networks
through the master stability function \cite{BP:2002,PC:1998}. To
apply this approach, the node dynamics in complex networks are
assumed to be identical, and then the theory of master stability
function provides a general mathematical framework to relate the
synchronizability of a network to the spectral properties of the
corresponding coupling matrix
\cite{BP:2002,SF,PLGK:2006,GWLWL:2008,Weighted}.

The above two approaches allows an analytical and therefore
relatively more complete understanding of synchronization in
complex networks. Nevertheless, in realistically physical or
biological situations the node dynamics can conceivably be much
more complicated, and in general be different from each other. The
key point is that the possible collective behaviors for coupled
nonidentical oscillators should be of the type of GS rather than
CS and PS. It is thus of great importance to investigate the
collective behaviors on complex networks where the local node
dynamics might be different from each other. In this paper, we
present a systematical study on GS phenomenon in various complex
networks. Our particular attention is paid to the development of
GS in networks. For typical complex networks, such as scale-free
networks, small-world networks, random networks and modular
networks, interestingly, we find that GS generally occurs in these
networks, regardless of whether the node dynamics are identical or
nonidentical. For coupled identical oscillators on complex
networks, a general bifurcation path toward the final global CS is
shown as: non-synchronization $\rightarrow$ PGS $\rightarrow$ GS
$\rightarrow$ PCS $\rightarrow$ CS. Here, PGS and PCS stand for
partial GS and partial CS, respectively. We further study how the
network topology competes with the local dynamics to affect the
development of GS on networks. It is shown that for linearly
coupled oscillators on heterogeneous networks, GS usually starts
from the hub nodes and then spreads to other nodes; while on
homogeneous networks, GS generally starts from the nodes whose
local dynamics are less chaotic in the sense that the largest
Lyapunov exponents have relatively smaller values. Moreover, we
analyze how different coupling strategies can affect the
development of GS on complex networks, a point which has not been
addressed before.

In processing this paper, we noticed a recent work which has
reported GS phenomenon in scale-free networks \cite{HHHH:2008}.
Compared with the results in Ref. \cite{HHHH:2008}, the present
work is different in the following aspects. Firstly, in our work
GS is systematically investigated on various complex networks,
including scale-free networks, random networks, small-world
networks and modular-networks; while in Ref. \cite{HHHH:2008}, GS
is mainly studied on a very special network, i.e., the scale-free
networks with tree-like structure. Secondly, the present work
characterizes the complete bifurcation path for coupled identical
oscillators on complex networks, i.e., from non-synchronization to
global CS, including the transition regimes; while in Ref.
\cite{HHHH:2008}, global CS has not be achieved. Thirdly, the
present work investigates the development of GS for coupled
nonidentical oscillators, either parametrically different or
physically different; while Ref. \cite{HHHH:2008} does not
consider this general setting where the occurrence of GS is
naturally justified. Finally, in the present work, we discuss the
effect induced by different coupling strategies on the development
of GS on network. Especially, we analyze and explain the different
observations regarding the GS evolution on networks in this work
and in Ref. \cite{HHHH:2008}. In fact, in many aspects the current
study deepens and widens the work in Ref. \cite{HHHH:2008}, and
can offer more thorough and comprehensive insight in understanding
GS phenomenon on complex networks.

This paper is organized as follows.  In Sec. II,  the methods to
characterize GS and CS on complex networks are described. In Sec.
III, GS of coupled identical oscillators is studied on complex
networks. The complete bifurcation path from non-synchronization
to global CS is studied and the dynamical features in the
transition regimes are characterized. In Sec. IV, GS of coupled
nonidentical oscillators is further studied on complex networks.
Our interest is focused on how the development of GS on networks
can be affected by network topology and the local dynamics. In
Sec. V, the effect of different coupling strategies on the
development of GS on complex networks is analyzed. Especially, GS
is observed in coupled system with hybrid oscillators even when
their local dynamics are physically different. Finally, a summary
ends this paper in Sec. VI.

\section{Approaches characterizing GS and CS on networks}
\label{sec:auxiliary_system_approach}

The auxiliary-system approach has been extensively used to detect
GS in two coupled chaotic systems \cite{ARS:1996}. Here, we can
extend it to detect GS on complex networks. The key observation is
that, for any given node in a network, the coupling from other
nodes can be regarded as a kind of ``driving.'' In particular, we
consider the following  linearly coupled identical oscillators on
a network:
\begin{equation} \label{eq:network_equation}
\dot{ \mathbf{x}}_i=\mathbf{F}_i(\mathbf{x}_i)-\varepsilon
                \sum_j a_{ij}(\mathbf{x}_i-\mathbf{x}_j),
\end{equation}
for $i = 1, \ldots, N$, where ${\bf x}_i$ denotes the dynamical
variables of node $i$, $\mathbf{F}_i(\mathbf{x}_i)$ is the local
vector field governing the evolution of ${\bf x}_i$ in the absence
of interactions with other nodes, $a_{ij}$ is the element of the
network adjacency matrix ${\bf A}$ ($a_{ij} = 1$ if there is a
link between node $i$ and node $j$, $a_{ij} = 0$ otherwise, and
$a_{ii} = 0$), and $\varepsilon$ is the coupling strength. To
apply the auxiliary-system approach, we consider a replica for
each oscillator in the original network:
\begin{equation} \label{eq:network_auxiliary_equation}
\dot{ \mathbf{x}}'_i=\mathbf{F}_i(\mathbf{x}'_i)-\varepsilon
                \sum_j a_{ij}(\mathbf{x}'_i-\mathbf{x}_j),
\end{equation}
for $i = 1, \ldots, N$. Note that the driving variable
$\mathbf{x}_j$ is identical for both Eqs.
(\ref{eq:network_equation}) and
(\ref{eq:network_auxiliary_equation}). If, for initial conditions
${\bf x}_i(0) \ne {\bf x}'_i(0)$, we have $|\mathbf{x}_i(t) -
\mathbf{x}'_i(t)| \rightarrow 0$ as $t\rightarrow\infty$, node $i
$ then is entrained in the sense that its dynamics is no longer
sensitive to initial conditions. In other words, there is GS
relation between ${\bf x}_i$ and ${\bf x}_j$ for $j = 1, \ldots,
N$. Numerically, we can examine the following local distance of GS
between a node and its auxiliary counterpart:
\begin{equation} \label{eq:local_syn_distance}
d(\varepsilon,i) =
\frac{1}{t_2-t_1}\sum^{t_2}_{t_1}|\mathbf{x}_i(t) -
\mathbf{x}'_i(t)|,
\end{equation}
where $t_1$ is chosen to be larger than the typical transient time
of the local dynamics ${\bf F}_i({\bf x}_i)$. For oscillators on
complex networks, GS may be gradually developed with the increase
of the coupling strength. To characterize the development of GS on
networks, we can define the distance of global GS as
$l_g(\varepsilon)= \langle d(\varepsilon,i) \rangle$. Here,
$\langle \cdot \rangle$ denotes the spatial average over all
nodes. If $l_g=0$, global GS has been achieved between any two
pair of oscillators on the whole network.

For coupled identical oscillators on complex networks, CS is
generally expected to take place as long as the network
connectivity is not too sparse. To characterize CS, we can define
$l_c(\varepsilon)$ as the distance of global CS, which measures
the distance between the dynamics of all oscillators and their
average, i.e.,
\begin{equation}\label{Distance_CS}
l_c(\varepsilon)=\frac{1}{t_2-t_1}\sum_{t_1}^{t_2} \langle |
\mathbf{x}- \langle \mathbf{x} \rangle | \rangle,
\end{equation}
where the meaning of $t_1$ and $t_2$ is the same as that in Eq.
(\ref{eq:local_syn_distance}). If $l_c=0$, global CS has been
achieved on the whole network.

For individual node dynamics, we choose the chaotic Lorenz
oscillator:
\begin{equation} \label{eq:Lorenz}
{\bf F}_i({\bf x}_i) =
[10(x_i-y_i),r_ix_i-y_i-x_iz_i,x_iy_i-(8/3)z_i]^T,
\end{equation}
where ${\bf x}_i \equiv (x_i,y_i,z_i)$ are the state variables of
the Lorenz oscillator. Note that in studying GS, the local
dynamics could be different from each other. This can be modelled
by setting different $r_i$ values for Lorenz oscillators in the
network. Without losing generality, the coupling between two nodes
in Eq. (\ref{eq:network_equation}) is through the $x$ variable.
For convenience, in the present work, we use two kinds of node
indices: $i_d$ and $i_r$. In the first index, we order the degrees
of the network so that $i_d = 1$ denotes the node with the largest
degree, $i_d=2$ is for the node with the second largest degree,
and so on. In the second index, we rank the Lorenz oscillators in
the network according to their $r_i$ values, i.e., $i_r = 1$
denotes the node with the largest parameter $r$, $i_r=2$ is for
the node with the second largest parameter $r$, and so on. These
two indices of nodes are used throughout the paper.

\section{Coupled identical oscillators: a general path
toward global CS}

\subsection{Path towards CS on scale-free networks}

\begin{figure} \begin{center}
\epsfig{figure=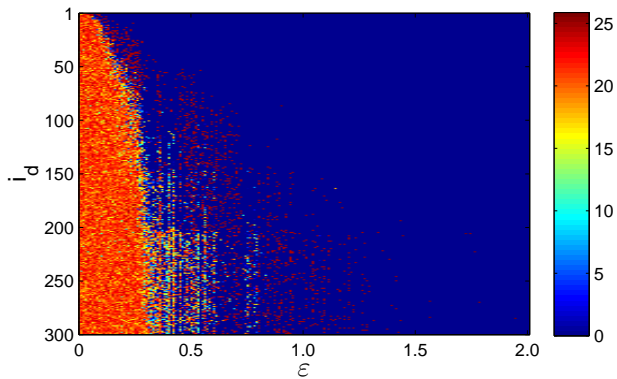,width=1.0\linewidth} \caption{(Color
online) Color-map of $d(\varepsilon,i)$ in the two-dimensional
parameter space $(\varepsilon,i)$, characterizing the development
of GS for  300 identical chaotic Lorenz  oscillators on a
scale-free network. The network is the BA model with $m_0=m=4$
\cite{BA:1999}. We see that with the increase of coupling
strength, GS can be achieved gradually from the hubs and then
spread to the rest of the network.   }
\end{center} \end{figure}

\begin{figure}[htbp] \begin{center}
\epsfig{figure=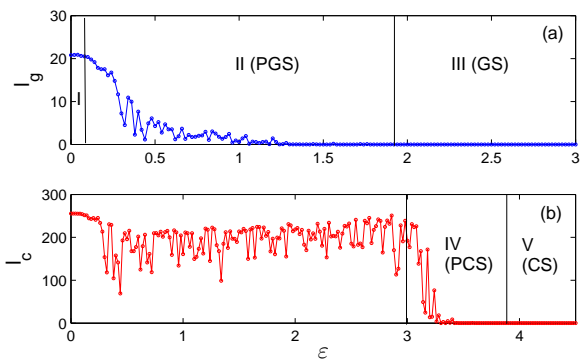,width=1.0\linewidth} \caption{(Color
online) Characterizing the path towards global CS on  scale-free
network. (a) The distance of global GS vs the coupling strength.
(b) The distance of global CS vs the coupling strength.
Numerically, if $l_g (l_c)<0.001$, global GS (CS) is regarded to
have been achieved. The notations of the regimes in this figure
are used throughout this paper.}
\end{center} \end{figure}

Previously, in the study of synchronization of two coupled chaotic
oscillators, it is found that CS generally takes place when the
coupled oscillators are identical, while GS usually occurs in
systems of coupled nonidentical oscillators. This observation is
accurate for most cases except for some specially designed
coupling strategies \cite{GKL:2006}. However, this generally
accepted view turns out to be not true when synchronization is
studied on complex networks. In this section, we investigate the
occurrence of GS for coupled identical chaotic oscillators on
various complex networks. Our particular interest is focused on
the complete bifurcations from initial non-synchronization towards
the final global CS with the increase of the coupling strength. By
applying auxiliary-system approach, interestingly, we find that GS
generally exists in coupled identical oscillators on various
complex networks, including scale-free networks, random networks,
small-world networks and modular networks.

As an example, we first study the bifurcation route towards global
CS for 300 coupled chaotic Lorenz oscillators on a scale-free
network. In this case, $r_i$=28 in Eq. (\ref{eq:Lorenz}) for all
nodes, i.e., all oscillators are identical. Intuitively, with the
increase of coupling, we expect that the coupled chaotic
oscillators will finally achieve global CS. The surprising finding
here is that before the system achieves the final global CS, there
exists another synchronization regime, i.e., the GS regime, which
usually occurs with much smaller coupling strength compared with
CS. To illustrate the GS phenomenon on scale-free network, we plot
the color-map of the distance matrix $d(\varepsilon,i_d)$ in Fig.
1. From Fig. 1, three distinct regimes of coupling can be
identified from the viewpoint of GS. When the coupling strength is
small, $d(\varepsilon, i_d)$ are greater than 0 for all nodes,
showing that the system is in the non-synchronous state. On the
other hand, when the coupling strength is large enough,
$d(\varepsilon, i_d)$ are 0 for all nodes, showing that all
oscillators have been entrained and the coupled system is in the
global GS state according to the auxiliary system approach
criterion. Between these two regimes, it is the regime of partial
GS, where part of the oscillators have been entrained but the
others have not. Obviously, this regime is a transition stage
between non-synchronous state and the global GS state.

To quantitatively characterize the GS regime for coupled chaotic
oscillators on scale-free network, in Fig. 2(a) we plot the
distance of global GS $l_g$ vs the coupling strength.  From this
figure, three regimes of coupling can numerically be identified.
On the other hand, for coupled identical oscillators,  CS is
generally expected to occur as long as the connections of the
network are dense enough. It is found that usually CS requires
larger coupling strength than GS. This implies that GS is a weaker
synchronization form compared with CS on complex networks, as in
the situation of coupled two-oscillator systems. To characterize
CS in the network, we plot the distance of global CS $l_c$ vs the
coupling strength in Fig. 2(b). We find that before CS is
achieved, there is also a transition stage, i.e., the partial CS
regime. Combining all the results in Fig. 2, we can identify five
dynamical regimes on the path towards global CS on the specific
scale-free network as follows. I: $\varepsilon \le 0.1$, the
non-synchronization regime; II: $ 0.1 < \varepsilon \le 1.8$, the
PGS regime; III: $ 1.8 < \varepsilon <3.0$, the global GS regime;
IV: $ 3.0 < \varepsilon <3.8 $, the PCS regime; V: $
\varepsilon>3.8 $, the global CS regime. These regimes thus
characterize the complete bifurcations from non-synchronization
state to global CS state for coupled identical oscillators on
scale-free networks.

\subsection{Dynamic and oscillatory transition to global
synchronization}\label{sec:PGS}

\begin{figure}
\begin{center}
\epsfig{figure=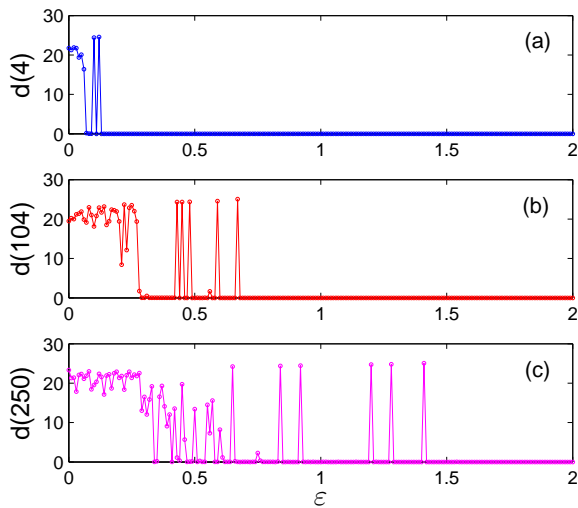,width=1.0\linewidth} \caption{(Color
online) Typical routes toward final GS for Lorenz oscillators in
scale-free network. From (a) to (c),  $d(\varepsilon,i_d)$ versus
$\varepsilon$ for $i_d $=4, 104, and 250, respectively. We observe
that nodes usually oscillate between the entrained state (GS) and
the unentrained state on the way toward final global GS state as
the coupling strength is increased. }
\end{center}
\end{figure}

\begin{figure}
\begin{center}
\epsfig{figure=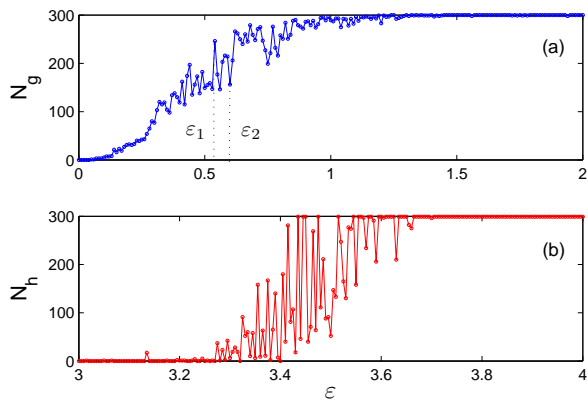,width=1.0\linewidth} \caption{(Color
online) Characterizing the dynamic and oscillatory nature in the
PGS and PCS regimes on a scale-free network. (a) The number of
nodes which have achieved GS vs the coupling strength. (b) The
number of nodes which have achieved CS with the hub (the node with
the largest degree) vs the coupling strength. }
\end{center}
\end{figure}

\begin{figure}
\begin{center}
\epsfig{figure=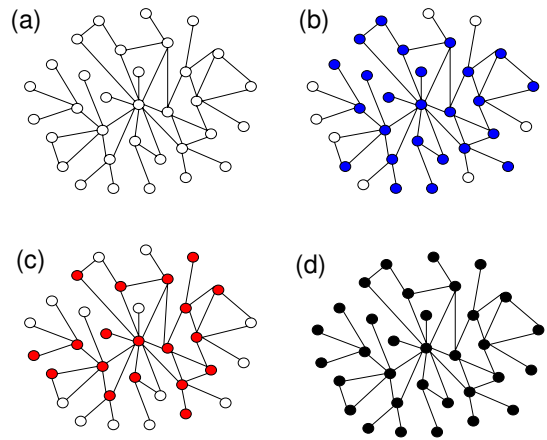,width=1.0\linewidth} \caption{(Color
online) A schematic visualization of the oscillatory path toward
global GS shown in Fig. 4(a). (a) Non-synchronous state at
$\varepsilon=0$. (b) PGS state at $\varepsilon_1=0.55$. About 80
percent nodes in the network (nodes with blue color) are
entrained. (b) PGS state at $\varepsilon_2=0.61$. About 50 percent
nodes in the network (nodes with red color) are entrained. (d)
Global GS state at $\varepsilon=3$. }
\end{center}
\end{figure}

In the above we have shown that there are two transition regimes,
i.e., the PGS regime and PCS regime, before the coupled system
achieves global GS and CS, respectively. In fact, these two
regimes are important to obtain information about how
synchronization is generated and spread on a network. As shown in
Fig. 1, with the increase of coupling strength, usually the hubs
of scale-free network become entrained first. Then more and more
nodes are entrained with further increase of the coupling
strength. At $\varepsilon \approx 0.3$, a giant cluster of
synchronous nodes in the sense of GS has been formed. The most
important dynamic feature in the PGS regime is that nodes may
oscillate between the entrained state and the unentrained state
with the increase of the coupling strength. As shown in Fig. 3, if
we focus on a specific node, we find that its path toward global
GS is generally oscillatory. Especially, for those nodes with
small degrees, the path could be extremely oscillatory as shown in
Fig. 3(c). As a consequence, in the PGS regime, the size of the
synchronous giant cluster does not increase monotonically, instead
it increases in an oscillatory manner with the increase of the
coupling strength, as shown in Fig. 4(a). Similar phenomenon has
been observed in the PCS regime. For example, in Fig. 4(b), we
plot the number of nodes which have achieved CS with the hub (with
the largest degree). This figure actually characterizes the
dynamical features for regime IV shown in Fig. 2(b). It is found
that the size of the synchronous cluster oscillates drastically on
the way toward global CS with the increase of the coupling
strength.

Recently, the path towards global PS in generalized Kuramoto model
has been studied in Ref. \cite{GMA:2007}. It is found that for
phase oscillators with the simplest local dynamics, once a phase
oscillator becomes frequency locked, it will never escape from the
synchronous cluster. Thus in the PS situation the size of the
synchronous cluster increases monotonically with the increase of
the coupling strength. Such a path toward global synchronization
can be called  as {\em static and monotonic}. In the present work,
we demonstrate a fundamentally different scenario towards global
synchronization for coupled chaotic oscillators on complex
networks. The most important feature of the path is that the
development and spreading of synchronization on networks is a
dynamic process during which oscillators may frequently oscillate
between the non-synchronous state and the synchronous state. This
implies that the size of the synchronous cluster might be
sensitive to the perturbation of the coupling strength in the
transition regimes. To visualize the dynamic feature of the
development of synchronization on networks,  a schematic figure
has been plotted in Fig. 5 for a scale-free network. Our results
suggest that in more realistic situations, the path towards global
synchronization on complex networks could be {\em dynamic and
oscillatory} rather than {\em static and monotonic}.

\begin{figure}[htbp] \begin{center}
\epsfig{figure=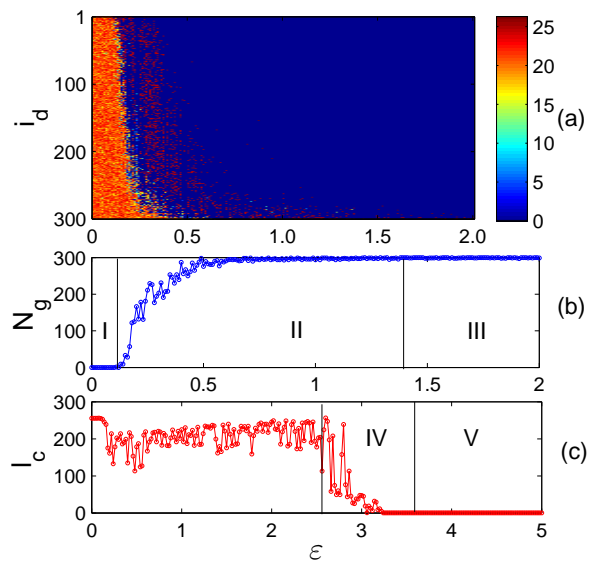,width=1\linewidth} \caption{(Color online)
Characterizing the development of GS and the path toward global CS
of 300 coupled identical Lorenz oscillators on a random network
with average degree 10. (a) The color-map of $d(\varepsilon,i)$ in
the two-dimensional parameter space $(\varepsilon,i)$.  (b) The
number of nodes which have achieved GS vs the coupling strength.
(c) The distance of global CS vs the coupling strength. }
\end{center} \end{figure}

\subsection{Path toward global CS on other  types of  complex networks}

In the above we have characterized the path towards global CS for
coupled identical chaotic oscillators on scale-free network, which
is characterized by a heterogeneous degree distribution. We have
also investigated synchronization path for coupled identical
chaotic oscillators on other typical complex networks.
Specifically, the networks studied include the following: (1) A
random network consisting of 300 nodes. The average degree is 10.
(2) A small-world network, which is obtained by rewinding 20 links
in a regular network consisting of 100 nodes and each node has 10
nearest neighbor connections \cite{WS:1998}. (3) A modular network
consisting of 100 nodes which is evenly divided into 5 modules.
Inside each module nodes are fully connected. Any two nodes in
different modules have probability $p=0.01$ to connect each other.
(4) As a special case of complex networks, a regular network
consisting of 100 nodes and each node has 10 nearest neighbor
connections.

The pathes towards global CS on these typical complex networks
have been characterized as for scale-free networks. Here, we only
show one example in the case of random network. In Fig. 6, the
bifurcation route to global CS for 300 identical Lorenz
oscillators on a random network is characterized. Comparing Fig.
6(a) with Fig. 1, we can see that the development of GS on random
network is different from the situation on scale-free network.
Mainly, most nodes are entrained at approximately the same
coupling strength. This is consistent with the fact that  random
network has approximately homogeneous degrees and thus no
extraordinary hubs exist as in scale-free networks. Although the
development of GS on different networks depends on the network
topologies, qualitatively, the path toward global GS is the same
as shown in Fig. 1 and Fig. 6. In fact, for other types of complex
networks, we have similar observations, as in the scale-free
network and random network. Based on these studies, we can draw
the following conclusions. (1) For coupled identical chaotic
oscillators, whether the network is homogeneous or heterogeneous,
a general bifurcation path towards global CS exists with the
increase of coupling strength, i.e, non-synchronization
$\rightarrow$ PGS $\rightarrow$ GS $\rightarrow$ PCS $\rightarrow$
CS. (2) Usually, global GS needs relatively smaller coupling
strength to achieve compared with global CS. Sometimes, the latter
requires ten times larger coupling strength to achieve than the
former. This shows that GS is a weaker synchronization form on
complex networks compared with CS, and is consistent with the
observation in coupled two-oscillator systems. (3) Typically, the
coupled system undergoes a {\em dynamic and oscillatory}
transition to achieve global synchronization, such as global GS or
global CS. During the PGS or PCS regime, oscillators may
frequently join in and escape from the giant synchronous cluster.
This leads to that the size of synchronous giant cluster increases
in an oscillatory manner with the increase of the coupling
strength. In some circumstances, the transition regimes cover
quite wide range of coupling strength, reflecting  that there
exists intense interaction among nodes in achieving global
synchronization on complex networks.

\section{Coupled parametrically different oscillators: topology vs local dynamics}

So far, most of the works on synchronization in complex networks
study the coupled identical oscillators, or coupled simple phase
oscillators of Kuramoto type. However, in real-world networks, the
interacting components are in principle different from each other.
For example, in neuron networks any two pair of neurons cannot be
exactly the same. This raises an important question: is it
possible for complex networks with different local dynamics
achieve GS, or certain coherence? If so, how network topologies
affect the GS? In this section, we will investigate
synchronization of coupled nonidentical oscillators on complex
networks. Our particular interest is to reveal how the development
of GS on networks is dominated by the interplay between  network
topology and the local dynamics.

As in the case of coupled system with two different oscillators,
we find that CS is generally forbidden for coupled nonidentical
oscillators on complex networks. Instead, we observe that GS
generally occurs in such coupled systems on networks. In our
study, the Lorenz oscillators, i.e., Eq. (\ref{eq:Lorenz}), are
made different on network by randomly setting parameter $r_i$ in
the interval $[28.0,30.0]$. By applying auxiliary system approach,
we can plot the color-map of the local synchronization distance
$d(\varepsilon,i_d)$ in the two-dimensional parameter plane
$(\varepsilon,i_d)$. As an example, Fig. 7 illustrates the
development of GS for Lorenz oscillators with parameter mismatches
on a scale-free network. It is shown that with the increase of
coupling strength, nodes in the network are entrained first from
the hubs and gradually spread to other nodes. When the coupling
strength is large enough, all nodes in the network are entrained
indicating global GS has been achieved. This situation is similar
to the case of coupled identical oscillators due to the fact that
the parameter mismatches here are small. Moreover, for coupled
nonidentical oscillators, the dynamical feature of the route
toward global GS is essentially the same as in the case of
identical oscillators.

\begin{figure}
\begin{center}
\epsfig{figure=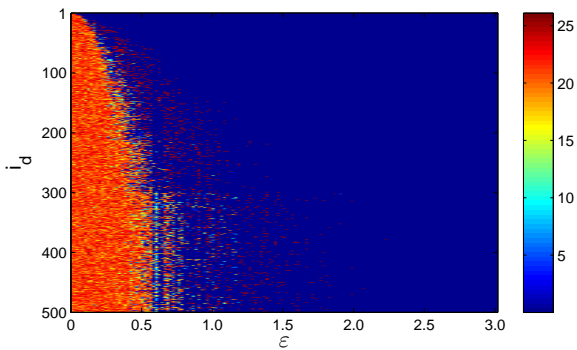,width=1.0\linewidth} \caption{(Color
online) Color-map of $d(\varepsilon,i_d)$ in the two-dimensional
parameter space $(\varepsilon,i)$, characterizing the development
of GS for 500 non-identical Lorenz chaotic oscillators on a
scale-free network. The network is the BA model with $m_0=m=3$
\cite{BA:1999}. The development of GS in the network is similar to
the situation shown in Fig. 1. }
\end{center}
\end{figure}

\begin{figure}
\begin{center}
\epsfig{figure=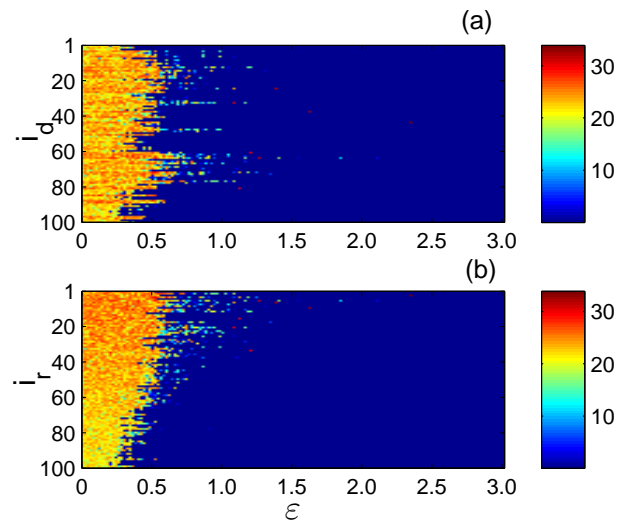,width=1.0\linewidth} \caption{(Color
online) Color-maps of $d(\varepsilon,i_d)$ and
$d(\varepsilon,i_r)$ in the two-dimensional parameter space
$(\varepsilon,i)$ for nonidentical Lorenz oscillators on a regular
network. The network has 100 nodes, and each node has 6 nearest
neighboring connections.  }
\end{center}
\end{figure}

\begin{figure}
\begin{center}
\epsfig{figure=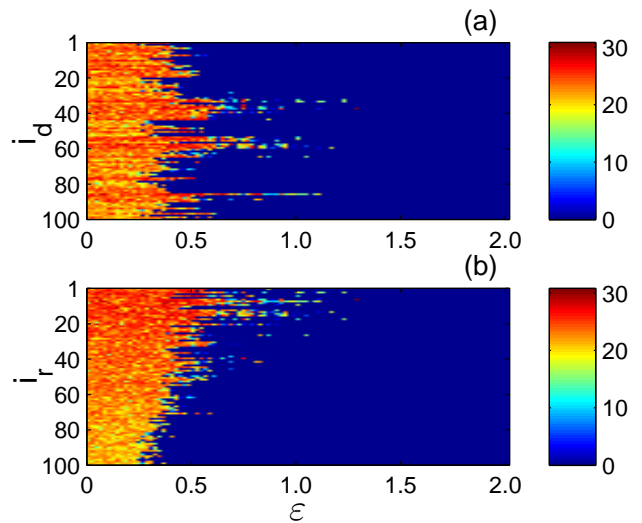,width=1.0\linewidth} \caption{(Color
online) Color-maps of $d(\varepsilon,i_d)$ and
$d(\varepsilon,i_r)$ in the two-dimensional parameter space
$(\varepsilon,i)$ for nonidentical Lorenz oscillators on a
small-world network. The network  is obtained by rewiring a small
part of connections on a regular network with 100 nodes
\cite{WS:1998}.}
\end{center}
\end{figure}

\begin{figure}
\begin{center}
\epsfig{figure=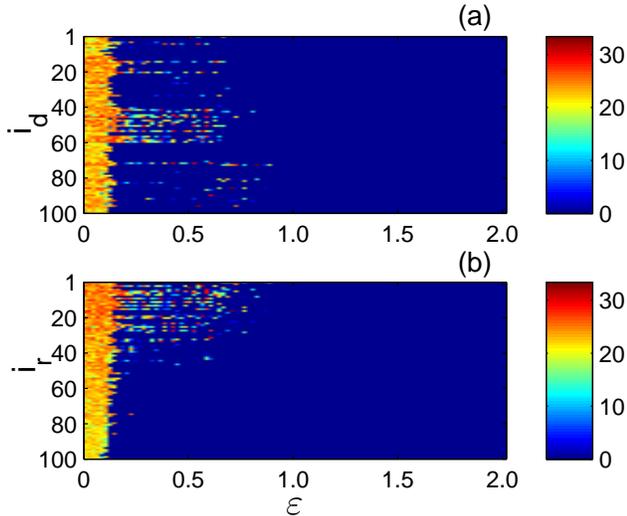,width=1.0\linewidth} \caption{(Color
online) Color-maps of $d(\varepsilon,i_d)$ and
$d(\varepsilon,i_r)$ in the two-dimensional parameter space
$(\varepsilon,i)$ for nonidentical Lorenz oscillators on a
clustered network. The network has 100 nodes which are evenly
divided into 5 modules. Nodes within each module are fully
connected. Any two nodes in different modules have probability
$\rho=0.005$ to connect each other. }
\end{center}
\end{figure}

The interplay between network dynamics and network topology is the
key to understand collective behaviors on complex networks.
Previously, many works have revealed how network topology affects
the network synchronizability under the setting of coupled
identical oscillators. For nonidentical oscillators, there are two
factors affecting the development of GS on complex networks, i.e.,
network topology and heterogeneity in the local dynamics. The
effect of heterogeneity in the local dynamics, if they are
chaotic, on the development of GS remains to be a challenging
problem. In the two preceding sections, we have shown that for
scale-free networks, GS typically starts from the hubs, and then
spreads to others nodes with relatively smaller degrees. In this
case, heterogeneity in the degree distribution appears to be the
main factor governing the development of GS. An interesting issue
is then that, for networks with homogeneous degree distributions,
such as regular, small-world, or certain modular networks, how
does the heterogeneity in the local dynamics affect GS? In the
following, we study this question numerically.

We first consider a regular network of $N=100$ nodes. Each node in
the network has $k_d=6$ connections to its nearest neighbors. As a
special case, this is an exactly homogeneous network with a
regular degree sequence. The local dynamics are those of Lorenz
chaotic oscillators with different parameter $r_i$ randomly
distributed in the interval [28, 38] for this example and for the
following two examples. Figure 8 shows the color-maps of the local
synchronization distances $d(\varepsilon,i_d)$ and
$d(\varepsilon,i_r)$, where $i_d$ and $i_r$ are the node indices
arranged according to the decreasing node degree and decreasing
values of parameter $r_i$, respectively. For the Lorenz oscillator
under the parameter setting in our study, we find that the larger
the value of parameter $r$, the larger the largest Lyapunov
exponent of the chaotic attractor. Thus, in Fig. 8, the index
$i_r$ actually corresponds to the decreasing value of the largest
Lyapunov exponent. Specifically, larger value of $i_r$ corresponds
to a small value of $r_i$ so that the corresponding local dynamics
is less chaotic in the sense that its largest Lyapunov exponent
has a relatively smaller value. Comparing Figs. 8(a) with 8(b), we
find that, when transient behaviors are disregarded, nodes whose
dynamics are less chaotic require smaller value of the coupling
strength to be entrained with other nodes in the network.
Entrainment of more chaotic nodes requires stronger coupling. Thus
the development of GS on regular network is dominated by the local
dynamics.

We next consider the occurrence and development of GS on
small-world networks. A representative example is shown in Fig. 9.
We see from Fig. 9(a) that there is no apparent synchronization
sequence of nodes according to the degree index $i_d$. However, as
can be seen from Fig. 9(b), GS starts from nodes with smaller
values of the largest Lyapunov exponent. This demonstrates that
for a small-world network, heterogeneity in the local dynamics
plays a dominant role in the development of GS, which is similar
to the situation in regular networks.

Lastly, we investigate GS in a type of modular network. A modular
network is characterized by a number of sparsely connected
subnetworks, each with dense internal connections. The modular
structure has been identified in biological, social, and certain
technological networks \cite{Clustered_Networks}. For such a
network, synchronization within each individual cluster can
usually be achieved readily due to the dense internal connections,
so the occurrence of global synchronization is particularly
interesting \cite{GWLL:2008}. An example of the development of GS
on modular network is shown in Fig. 10. We see that global GS can
be achieved despite the sparse inter-cluster connections. An
interesting phenomenon is that, the development of GS does not
seem to strongly depend on $i_d$ or $i_r$. Nevertheless,
comparatively, it still can be found that nodes with less chaotic
local dynamics are easier to be entrained with smaller coupling
strength. This can be explained by noting that the degree
distribution is approximately homogeneous, as a result of the
sparse inter-cluster connectivity. In addition, nodes with
different values of $r$ are uniformly distributed in different
synchronous clusters. As a result, the effect of heterogeneity in
the local dynamics is suppressed to some extent.

Based on the above studies, we can conclude that the development
of GS on complex networks is the result of interaction between
network topology and local dynamics when they both have
heterogeneity. For heterogenous networks, such as scale-free
networks, the network topology plays a leading role; while for
approximately homogeneous networks, such as small-world networks
and modular networks, the local dynamics is the dominant factor
affecting the development and spreading of GS on networks.

\section{Effect of different coupling strategies }

\subsection{The linear coupling and the normalized coupling}

We have seen that the hub nodes behave as  ``seeds'' to develop GS
in scale-free networks. The phenomenon that synchronization starts
from the hub nodes in a heterogeneous network is in fact quite
general, which holds for both GS and CS. Qualitatively, this can
be understood by regarding the interactions among nodes as a kind
of ``noisy'' driving within the network. In particular, since the
dynamics of each individual node is chaotic, the coupling term in
Eq. (\ref{eq:network_equation}) for node $i$ is also chaotic,
which can be regarded effectively as correlated noisy driving. It
has been known that correlated noise can induce and/or enhance
synchronization \cite{Common_noise_syn}, PS \cite{ZK:2002}, and GS
as well \cite{GLL:2006}. That is, correlated noise of larger
amplitude can cause oscillators under such driving to be better
synchronized. For a heterogeneous network, hub nodes have
significantly more links than other nodes. Since the drivings from
different nodes are not entirely uncorrelated, the equivalent
driving force to the hub nodes are generally stronger,
facilitating their synchronization with other nodes in the
network.

However, in a recent publication studying GS on networks
\cite{HHHH:2008}, it is reported that only for a kind of special
scale-free network with tree-like structure, GS is observed to
develop from the hubs and then gradually spread to other nodes in
the network; while for usual scale-free networks, the
heterogeneity of network seems to have little effect on the
development of GS in the network. In attempting to find the reason
that leads to different observations in the present work and Ref.
\cite{HHHH:2008}, we notice that a different coupling strategy is
used in Ref. \cite{HHHH:2008}. Take a coupled one-dimensional map
network as an example, the coupled system in Ref. \cite{HHHH:2008}
is:
\begin{equation} \label{eq:Hung}
x_i^{n+1}=(1-\varepsilon)f(x_i^n)+\frac{\varepsilon}{k_i}
                \sum_j a_{ij}f(x_j^n),
\end{equation}
where $x_i^n$ is the state variable of node $i$ at time step $n$,
$f(x)$ is the local map, and $k_i$ is the degree of node $i$. This
coupling scheme with each node in the network driven by the local
mean field is essentially the same as the following normalized
coupling strategy:
\begin{equation} \label{eq:network_equation2}
\dot{\mathbf{x}}_i=\mathbf{F}_i(\mathbf{x}_i)-\frac{\varepsilon}{k_i}
                \sum_j a_{ij}(\mathbf{x}_i-\mathbf{x}_j),
\end{equation}
where notations are the same as that in Eq.
(\ref{eq:network_equation}) and Eq. (\ref{eq:Hung}). In this
coupling scheme, the coupling strength for each node $i$ is
normalized by its degree. Therefore,  the effective coupling
strength each node received from its network neighbors is of the
same order of magnitude, regardless it is a hub or a node with
very small degree. Obviously, due to the normalization, the effect
of network topology on the development of GS has been almost
suppressed. As a consequence, there should be no distinct
difference among the coupling strength when nodes  in the network
achieve GS. Quantitatively, we can find that in order to obtain
the same effective coupling strength, system
(\ref{eq:network_equation2}) should require $\langle k \rangle$
times larger coupling strength than system
(\ref{eq:network_equation}). Here,  $\langle k \rangle$ denotes
the average degree of the network. Thus on an average, it can be
expected that system (\ref{eq:network_equation2}) needs $\langle k
\rangle$ times larger coupling strength to achieve GS in network
compared with system (\ref{eq:network_equation}).

The above heuristical considerations can be   further analytically
explained. By examining the coupled system
(\ref{eq:network_equation}) and its corresponding auxiliary system
 (\ref{eq:network_auxiliary_equation}). Letting $\Delta {\bf x}_i
= {\bf x}'_i - {\bf x}_i$ and subtracting Eq.
(\ref{eq:network_equation}) from Eq.
(\ref{eq:network_auxiliary_equation}), we obtain
\begin{eqnarray} \label{eq:Delta_x_evolution}
\dot{\Delta {\bf x}_i} & = &  {\bf F}_i({\bf x}'_i) - {\bf
F}_i({\bf x}_i) - \varepsilon \sum_j a_{ij} \Delta {\bf x}_i \\
\nonumber & \approx &  [{\bf DF}({\bf x}_i) - \varepsilon k_i {\bf
I}]\cdot \Delta {\bf x}_i,
\end{eqnarray}
where ${\bf I}$ is the $M\times M$ identity matrix with $M$ the
dimension of local dynamics. We see that, for node $i$, the
effective coupling strength is proportional to its degree $k_i$.
For a fixed value of $\varepsilon$, the coupling between a hub
node and its counterpart in the auxiliary system can be
significantly larger than the coupling for nodes with smaller
degree, leading to ``earlier'' synchronization between the hub
nodes in the original and in the auxiliary systems. Therefore, for
usual linear coupled oscillator system, the general observation is
that, in a complex network with heterogeneous degree distribution,
the set of hub nodes provides a skeleton around which
synchronization is developed. The above analysis is also suitable
for coupled system (\ref{eq:network_equation2}). In this case,
there will be no factor $k_i$ as in Eq.
(\ref{eq:Delta_x_evolution}). This implies that the effect of
network topology on the development of GS in networks is
eliminated.

\begin{figure}
\begin{center}
\epsfig{figure=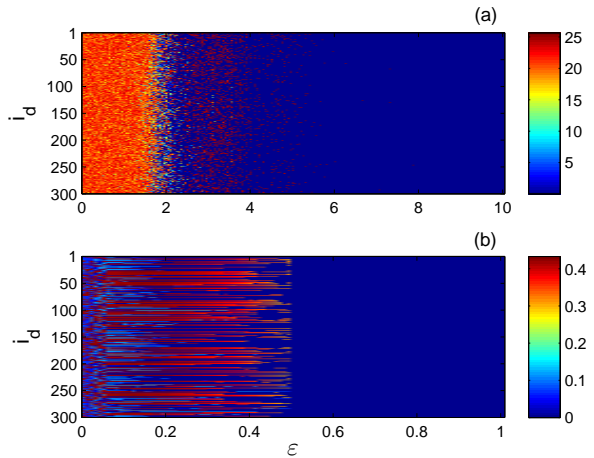,width=1.0\linewidth} \caption{(Color
online) Color-maps of $d(\varepsilon,i_d)$ in the two-dimensional
parameter space $(\varepsilon,i)$, characterizing the development
of GS for 300 identical Lorenz chaotic oscillators (a), and 300
identical Logistic maps (b) on a scale-free network. The network
is the same as in Fig. 1. Compared with Fig. 1, the effect caused
by the normalized coupling strategy is obvious. }
\end{center}
\end{figure}

In the following of this section, we further present two examples
where the oscillators in a scale-free network are interacted with
the normalized coupling strategy. The first example is coupled
Lorenz oscillators described by Eq. (\ref{eq:network_equation2}),
and the second example is the following coupled Logistic maps:
\begin{equation} \label{eq:Logistic_map}
x_i^{n+1}=f(x_i^n)-\frac{\varepsilon}{k_i}
                \sum_j a_{ij}[f(x_i^n)-f(x_j^n)].
\end{equation}
For both cases, the local dynamics are identical, i.e., $r_i$=28
for all Lorenz oscillators and $f(x)=4x(1-x)$ for all Logistic
maps. Figure 11 illustrates the development of GS for the above
two systems on a scale-free network. The effect of different
coupling strategies on the development of GS  can be verified by
comparing Fig. 11 with Fig. 1. For system
(\ref{eq:network_equation}), GS first occurs on the hubs and then
gradually spread to other nodes; while for systems
(\ref{eq:network_equation2}) and (\ref{eq:Logistic_map}), GS
almost simultaneously takes place on the hubs and the other nodes.
Actually, there is no significant difference among all nodes on
networks. By examining Fig. 1 and Fig. 11(a), it is found that the
giant GS cluster forms at $\varepsilon \approx 0.5$ and
$\varepsilon \approx 2$,  respectively, in the two cases. In
particular, the latter is approximately four times larger than the
former. Note that the scale-free networks we used in Fig. 1 and
Fig. 11(a) are BA model with $m_0=m=4$, so the average node degree
$\langle k \rangle $=4. This is consistent with the theoretical
analysis above.

\subsection{Coupled hybrid oscillators: competition of local dynamics }

\begin{figure}
\begin{center}
\epsfig{figure=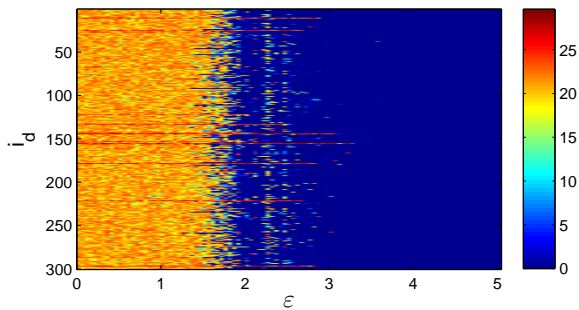,width=1.0\linewidth} \caption{(Color
online) Color-map of $d(\varepsilon,i_d)$ in the two-dimensional
parameter space $(\varepsilon,i)$, characterizing 300 hybrid
Lorenz chaotic oscillators on a scale-free network. The network is
the BA model with $m_0=m=3$ \cite{BA:1999}.  }
\end{center}
\end{figure}

\begin{figure}
\begin{center}
\epsfig{figure=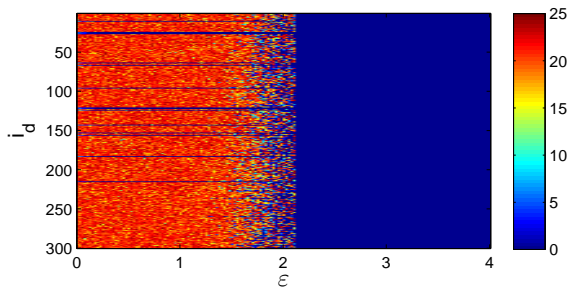,width=1.0\linewidth} \caption{(Color
online) Color-map of $d(\varepsilon,i_d)$ in the two-dimensional
parameter space $(\varepsilon,i)$, characterizing 300 hybrid
Lorenz and Rossler oscillators on a scale-free network. The
network is the BA model with $m_0=m=3$ \cite{BA:1999}. }
\end{center}
\end{figure}

In Sec. IV, we have studied how network topology competes with
local dynamics to dominate the development of GS in complex
networks when there are heterogeneity in both network topology and
local dynamics. In fact, for coupled nonidentical oscillators on
complex networks, competition may also occur among local dynamics
if the parameter mismatches of local dynamics are large enough, or
the local dynamics are physically different. By adopting the
normalized coupling strategy, we can conveniently investigate this
issue. In the following, we present two examples showing how GS is
developed for coupled hybrid oscillator system in networks. By
hybrid we mean that the oscillators in the network can be
classified into different types which are parametrically different
or physically different. In the first example, we consider a
system coupled with hybrid Lorenz oscillators. To be specific, we
arbitrarily select 5\% Lorenz oscillators in the network and make
their parameter $r_i$ be randomly distributed in the interval [30,
40]. For the rest oscillators, they are the same with $r_i$=28,
which are significantly smaller than that of the 5\% oscillators.
In Fig. 12, the development of GS for such a hybrid system is
illustrated on a scale-free network. It is seen that most
oscillators in the network are entrained at $\varepsilon\approx2$.
However, a small number of oscillators require significantly
larger coupling strength to be entrained. A careful examination of
the locations of these nodes reveals that they just correspond to
the 5\% oscillators with larger $r$ values. In the second example,
the coupled hybrid system consists of two kinds of oscillators
which have different dynamical equations. Similar to the first
example, 5\% nodes are randomly selected to be the identical
Rossler oscillators, while the rest 95\% percent nodes are the
identical Lorenz oscillators with $r_i$=28. The dynamical
equations of the Rossler oscillator read:
\begin{equation} \label{eq:Rossler}
{\bf F}({\bf x}_i) = [-(y_i+z_i),x_i+0.2y_i,0.2+z_i(x_i-5.7)]^T,
\end{equation}
where ${\bf x}_i \equiv (x_i,y_i,z_i)$ are the state variables of
the Rossler oscillator. In Fig. 13, we plot the local distance of
GS vs the coupling strength for all nodes in a scale-free network.
It is clearly shown GS can be achieved for such a coupled hybrid
system. The development of GS in network has two distinct stages:
the Rossler oscillators are much easier to be entrained compared
with the Lorenz oscillators.  The latter needs quite larger
coupling strength to achieve GS. Note that the largest Lyapunov
exponent of the Lorenz oscillator is proportional to its $r$
value, and  the Rossler oscillator has much smaller Lyapunov
exponent than that of the Lorenz oscillator, we can conclude from
the above two examples that GS usually develops from the nodes in
the network where the local dynamics are less chaotic in the sense
that the largest Lyapunov exponents of the local dynamics have
relatively smaller values. Although our numerical simulations are
carried out on a scale-free network, we believe this conclusion
can still hold for other network topologies when the normalized
coupling strategy is applied.

\section{Summary}

In summary, we have investigated the occurrence and development of
GS on various complex networks, including scale-free networks,
small-world networks, random networks and modular networks. It is
shown that GS generally takes place in such networks for both
coupled identical oscillators and nonidentical oscillators. For
coupled identical oscillators, we reveal a general bifurcation
path to global CS, i.e., non-synchronization $\rightarrow$ PGS
$\rightarrow$ GS $\rightarrow$ PCS $\rightarrow$ CS. Prior to the
onset of global GS (CS), there exists a coupling regime of partial
GS (CS) characterized by the occurrence of a cluster of
synchronous nodes. In the transient regimes, the size of the
synchronous cluster usually oscillates with the increase of
coupling strength, showing a dynamic scenario when the coupled
system is approaching global synchronization. We find that the
development of GS on complex networks depends on both network
topology and local dynamics. Moreover, the specific coupling
strategy also plays an important role during the evolution of
synchronization. We show that under the usual linear dissipative
coupling scheme, for heterogeneous networks, GS generally starts
from a small number of hub nodes, and then spread to the rest
nodes in the network; while for homogeneous networks, GS usually
starts from the nodes whose local dynamics are less chaotic in the
sense that the largest Lyapunov exponents have relatively smaller
values. We further show that the effect of network topology on the
development of GS can be suppressed if the coupling strengths of
nodes in the network are normalized. Under such coupling scheme,
the development of GS is essentially determined by the competition
among different local dynamics. We demonstrate that GS can also
occur in coupled systems with hybrid oscillators on complex
networks, and the development of GS has distinct stages
corresponding to physically different local dynamics. We note
that, while there are extensive works on synchronization in
complex networks, prior to this work the bifurcation path toward
global CS via GS, the dynamical features of the development of GS,
especially the interplay between the network topology and local
dynamics on typical complex networks have not been investigated.
Our work reveals that  on complex networks,  coupled oscillators
may present fundamentally different synchronization regimes which
deserves further study.

\section{Acknowledgment}

This work is supported by Temasek Laboratories at National
University of Singapore.

\end{document}